\documentclass[dvipdfm]{elsart}

\usepackage{amssymb}
\usepackage{graphicx}

\newcommand{\sqrtsnn}{\sqrt{s_{_{NN}}}}
\def\mean#1{\ensuremath{\left<#1\right>}}

\newcommand{\gaga}{\gamma\,\gamma}

\newcommand{\gA}{\gamma\,A}

\def\ee{\mbox{$e^+e^-$}}

\def\tab#1{{Table~\ref{#1}}}
\def\fig#1{{Figure~\ref{#1}}}
\providecommand{\jpsi}{J/\psi}

\providecommand{\mean}[1]{\ensuremath{\left<#1\right>}}

\begin{document}

% location of figures
%\newcommand \figs{/phenix/WWW/p/info/ppg/07X/figures/eps}
%\newcommand \figs{/phenix/WWW/p/draft/silvermy/run4/upc/PLB/draft/figures/eps}
%\newcommand \figs{./figures/eps}

%\graphicspath{{\figs/}}

\begin{frontmatter}


\title{Photoproduction of $\jpsi$ and of high mass $\ee$ \\
	in ultra-peripheral Au+Au collisions \\
	at $\sqrtsnn$~=~200~GeV}

\author{S.~Afanasiev,$^{q}$}
\author{C.~Aidala,$^{g}$}
\author{N.N.~Ajitanand,$^{aq}$}
\author{Y.~Akiba,$^{ak,al}$}
\author{J.~Alexander,$^{aq}$}
\author{A.~Al-Jamel,$^{ag}$}
\author{K.~Aoki,$^{w,ak}$}
\author{L.~Aphecetche,$^{as}$}
\author{R.~Armendariz,$^{ag}$}
\author{S.H.~Aronson,$^{c}$}
% ...
\\author{(PHENIX Collaboration)}
% ...
\address[c]{Brookhaven National Laboratory, Upton, NY 11973-5000, U.S.}
% ...
\address[g]{Columbia University, New York, NY 10027 and Nevis Laboratories, Irvington, NY 10533, U.S.}
% ...
\address[q]{Joint Institute for Nuclear Research, 141980 Dubna, Moscow Region, Russia}
% ...
\address[w]{Kyoto University, Kyoto 606-8502, Japan}
% ...
\address[ag]{New Mexico State University, Las Cruces, NM 88003, U.S.}
% ...
\address[ak]{RIKEN Nishina Center for Accelerator-Based Science, Wako, Saitama 351-0198, JAPAN}
\address[al]{RIKEN BNL Research Center, Brookhaven National Laboratory, Upton, NY 11973-5000, U.S.}
% ...
\address[aq]{Chemistry Department, Stony Brook University, Stony Brook, SUNY, NY 11794-3400, U.S.}
% ...
\address[as]{SUBATECH (Ecole des Mines de Nantes, CNRS-IN2P3, Universit{\'e} de Nantes) BP 20722 - 44307, Nantes, France}
% ...
\thanks[spokes]{PHENIX Spokesperson: jacak@skipper.physics.sunysb.edu}
\thanks[deceased]{Deceased}

\begin{abstract}

We present the first measurement of photoproduction of $\jpsi$ and 
of two-photon production of high-mass $e^+e^-$--pairs in 
electromagnetic (or ultra-peripheral) nucleus-nucleus interactions, 
using Au+Au data at $\sqrtsnn$~=~200~GeV. The events are tagged 
with forward neutrons emitted following Coulomb excitation of one 
or both $Au^\star$ nuclei. The event sample consists of 28 events 
with $m_{e^+e^-}>$~2~GeV/$c^2$ with zero like-sign background. The 
measured cross sections at midrapidity of $d \sigma/dy \, (\jpsi + 
Xn, y=0) =$ 76 $\pm$ 33 (stat) $\pm$ 11 (syst) $\mu$b and $d^2 
\sigma/dm dy \, (e^+e^- + Xn, y=0) =$ 86~$\pm$~23~(stat)~$\pm$ 16 
(syst) $\mu$b/(GeV/c$^2$) for $m_{e^+e^-} \in 
\mbox{[2.0,2.8]}$~GeV/c$^2$ are consistent with various theoretical 
predictions.

\end{abstract}

\end{frontmatter}

% main text

%%%%%%%%%%%%%%%%%%%%%%%%%%%%%%%%%%%%%%%%%%%%%%%%%%%%%% \input{introduction}
\section{Introduction}
\label{section:introduction}

\setcounter{section}{1}\setcounter{equation}{0}

The idea to use the strong electromagnetic fields present in high-energy 
nucleus-nucleus collisions to study photoproduction at hadron colliders has 
attracted growing interest in recent years, 
see~\cite{Baur:2001jj,Bertulani:2005ru,Baltz:2007kq} for reviews. 
Electromagnetic interactions can be studied without background from hadronic 
processes in ultra-peripheral collisions (UPC) without nuclear overlap, i.e. impact 
parameters larger than the sum of the nuclear radii. This study focuses on 
the measurement of exclusively produced high-mass $e^+ e^-$ pairs at 
midrapidity in Au+Au collisions at $\sqrtsnn$~=~200~GeV, $Au+Au \rightarrow 
Au+Au + e^+ e^-$. The results have been obtained with the PHENIX 
detector~\cite{Adcox:2003zm} at the BNL Relativistic Heavy Ion Collider 
(RHIC).

The electromagnetic field of a relativistic particle can be represented by a 
spectrum of equivalent photons. This is the Weizs{\"a}cker-Williams method of 
virtual quanta~\cite{vonWeizsacker:1934sx,Williams:1934ad}. The number of 
photons in the spectrum is proportional to $Z^2$, where $Z$ is the charge of 
the particle, and the equivalent two-photon luminosity is thus proportional 
to $Z^4$. The strong dependence on $Z$ favours the use of heavy ions for 
studying two-photon and photonuclear processes. The virtualities of the 
equivalent photons when the field couples coherently to the entire nucleus 
are restricted by the nuclear form factor to 
$Q^2=\left(\omega^2/\gamma^2+q_\perp^2\right)\lesssim (\hbar/R_A)^2$. Here, 
$\omega$ and $q_\perp$ are the photon energy and transverse momentum, 
respectively, $R_A$ is the nuclear radius and $\gamma$ the Lorentz factor of 
the beam. At RHIC energies, $\gamma$ = 108 and the maximum photon energy in 
the center-of-mass system is of the order of $\omega_{\rm max}\sim$~3~GeV 
corresponding to maximum photon-nucleon and two-photon center-of-mass 
energies of $W^{\rm max}_{\gamma N} \sim$~34~GeV and $W^{\rm max}_{\gamma \gamma} 
\sim$~6~GeV.

\begin{figure}[tb]
\begin{center}
\includegraphics[width=1.0\linewidth]{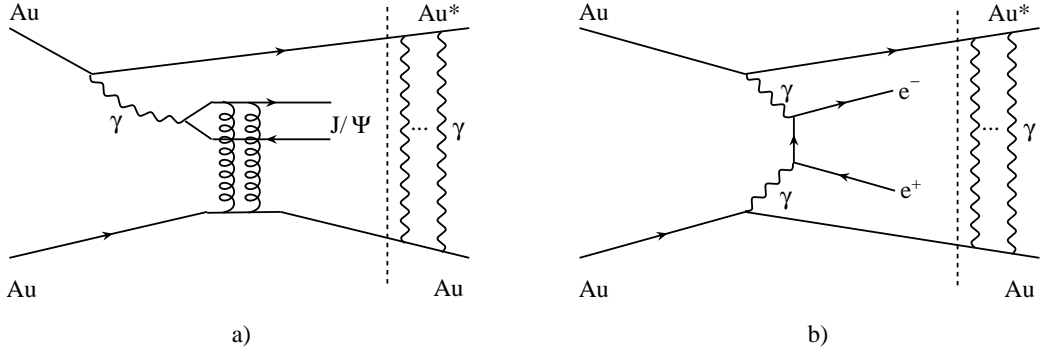}
\end{center}
\caption{Lowest order Feynman diagrams for exclusive photoproduction of (a) 
$\jpsi$ and (b) dielectrons, in ultra-peripheral Au+Au collisions. The 
photons to the right of the dashed line are soft photons that may excite the 
nuclei but do not lead to particle production in the central rapidity region. 
Both diagrams contain at least one photon and occur when the nuclei are 
separated by impact parameters larger than the sum of the nuclear radii.} 
\label{fig:diag_gg_gA} \end{figure}

The exclusive production of an $e^+ e^-$ pair can proceed either through a 
purely electromagnetic process (a two-photon interaction to leading order) or 
through coherent photonuclear production of a vector meson, which decays into 
an electron pair. Exclusive photoproduction of vector mesons is usually 
thought of as proceeding via Pomeron-exchange, the perturbative-QCD 
equivalent of which is the exchange of two gluons or a gluon ladder. The 
Feynman diagrams for the two leading order processes are shown in 
\fig{fig:diag_gg_gA}. The two-gluon picture is applicable to production of 
heavy vector mesons, such as the $\jpsi$, and to production of lighter mesons 
at high momentum transfers~\cite{Mohrdieck:2002cf}. The $\jpsi$ production 
cross section is consequently a good probe of the proton~\cite{Ryskin:1995hz} 
and nuclear gluon distribution, $G_A(x,Q^2)$, as well as of vector-meson 
dynamics in nuclear matter~\cite{Frankfurt:1995jw,Frankfurt:2001db}. For 
$\jpsi$-production, the coverage of the PHENIX central tracking arm, $-0.35 < 
\eta < 0.35$ corresponds to a range in the photon-nucleon center-of-mass 
energy between $21 < W_{\gamma N} <$~30~GeV, with a mean $\mean{W_{\gamma 
N}}$~=~24~GeV. This corresponds to photon energies in the rest frame of the 
target nucleus of $240 < E_{\gamma} <$~480~GeV, with 
$\mean{E_{\gamma}}$~=~300~GeV. Mid-rapidity photoproduction of $\jpsi$ probes 
nuclear Bjorken-$x$ values of $x = m_{\jpsi}^2/W_{\gamma A}^2 \approx 1.5 
\cdot 10^{-2}$~\cite{Frankfurt:2001db}, where the nuclear gluon density is 
partially depleted by ``shadowing'' effects~\cite{Armesto:2006ph} compared to 
the proton.

The strong fields associated with heavy ions at high energies lead to large 
probabilities for exchanging additional soft photons in the same event. Most 
of these photons have too low energy to produce particles, but they can 
excite the interacting nuclei. The dominating excitation is to a Giant-Dipole 
Resonance (GDR) with energies $\mathcal{O}(10$~MeV$)$, which decays by 
emitting neutrons at very forward rapidities, providing a very useful means 
to trigger on UPCs with Zero-Degree Calorimeters (ZDC). 
The probability for having a Coulomb excitation leading to 
emission of neutrons in at least one direction in coincidence with coherent 
$\jpsi$ production is 55\% $\pm$ 6\%~\cite{Baltz:2002pp}. The probabilities 
for exchanging one or several photons factorise, i.e. the Coulomb tagging 
does not introduce any bias in the extraction of exclusive $\jpsi$ 
photoproduction cross sections from these events~\cite{Baur:2003ar}. The soft 
photons leading to moderate nuclear excitation are indicated to the right of 
the dashed line in \fig{fig:diag_gg_gA}. Incoherent (or quasi-elastic) 
vector-meson photoproduction can also proceed via the interaction of the 
exchanged photon with a {\it single} nucleon in the nucleus. In that case, 
$\jpsi$ photoproduction is always accompanied by nuclear breakup and emission 
of nucleons in the forward direction~\cite{Strikman:2005ze}.

%|%%%%%%%%%%%%%%%%%%%%%%%%%%%%%%%%%%%%%%%%%%%%%%%%%%%%%%% Fig_1

Photoproduction of vector mesons has been studied with lepton beams first in 
the 60s~\cite{lanz:65,bau:78} and more recently at the electron-proton 
collider HERA~\cite{Chekanov:2002xi,Aktas:2005xu}. Measurements of 
photonuclear production of $\rho$ mesons~\cite{Adler:2002sc,Abelev:2007nb}, 
as well as of two-photon production of {\it low-mass} $e^+ e^-$ 
pairs~\cite{Adams:2004rz} in heavy ion interactions have been performed by 
the STAR collaboration. The CDF collaboration has studied two-photon 
production of $e^+ e^-$ pairs~\cite{Abulencia:2006nb} and exclusive 
production of $\mu^+ \mu^-$ pairs~\cite{Aaltonen:2009kg} in $p \overline{p}$ 
collisions at the Tevatron.  The PHENIX analysis presented here is the first 
on heavy final states in ultra-peripheral nucleus-nucleus collisions. It 
supersedes a preliminary study presented earlier~\cite{d'Enterria:2006ep}.  
The cross section for $\jpsi$ and $e^+ e^-$ photoproduction are compared with 
various theoretical 
calculations~\cite{Frankfurt:2001db,Strikman:2005ze,Klein:1999qj,Goncalves:2005sn,Ivanov:2007ms}.

\section{Experimental setup}
\label{section:experimental}

The data presented here were collected with the PHENIX detector at RHIC 
during the 2004 high-luminosity Au+Au run at $\sqrtsnn$~=~200~GeV. 
The PHENIX detector~\cite{Adcox:2003zm}, is a versatile detector designed to 
study the properties of strongly interacting matter at extreme temperatures 
and energy densities present in central heavy ion collisions. The current 
analysis demonstrates its capabilities to also study ultra-peripheral 
collisions, which have a very different event topology. The PHENIX central 
tracking system~\cite{Adcox:2003_ca} consists of two arms, each covering 
$| \eta | <$~0.35 and $\Delta \phi = \pi/2$, equipped with multi-layer drift 
chambers (DC) followed by multi-wire proportional chambers (PC) with 
pixel-pad readout. The tracking arms also have Ring-Imaging-\v{C}erenkov 
(RICH, with CO$_{2}$ gas radiator) detectors~\cite{Aizawa:2003} and 
electromagnetic calorimeters (EMCal)~\cite{Aphecetche:2003zr} for electron 
and positron identification. The PHENIX EMCal consists of six sectors of 
lead-scintillator sandwich calorimeter (PbSc, 15552 individual towers with 
5.54 cm $\times$ 5.54 cm $\times$ 37.5 cm, 18$X_0$) and two sectors of 
lead-glass \v{C}erenkov calorimeter (PbGl, 9216 modules with 4 cm $\times$ 4 
cm $\times$ 40 cm, 14.4$X_0$), at a radial distance of $\sim$5\,m from the 
beam line.

The ultra-peripheral Au+Au events were tagged by neutron detection 
at small forward angles in the ZDC. The 
ZDC's~\cite{Adler:2000bd,Chiu:2001ij} are hadronic calorimeters placed 
18~m up- and down-stream of the interaction point that measure the energy of 
the neutrons coming from the Au$^\star$ Coulomb dissociation with $\sim$20\% 
energy resolution and cover $|\theta|<$ 2 mrad, which is a very forward 
region\footnote{Much larger than the crossing angle of Au beams at the 
PHENIX interaction point (0.2 mrad).}.

The events used in this analysis were collected with the UPC trigger set up 
for the first time in PHENIX during the 2004 run with the following 
characteristics:

\begin{enumerate}

\item A veto on coincident signals in both Beam-Beam Counters (BBC, covering 
$3.0<|\eta|<3.9$ and full azimuth) selects exclusive-type events 
characterised by a large rapidity gap on either side of the central arm.

\item The EMCal-Trigger (ERT) with a 2$\times$2 tile threshold at 0.8~GeV. 
The trigger is set if the analog sum of the energy deposit in a 2$\times$2 
tile of calorimeter towers is above threshold (0.8~GeV).  The efficiency for 
triggering at least one of the two high-energy $e^\pm$ coming from the 
pair\footnote{ For instance, for the low-$p_{\rm T}$ photoproduced $\jpsi$, the 
$e^\pm$ decays have $E\approx 0.5\,m_{\jpsi}\approx$~1.6~GeV. } is estimated 
to be $\varepsilon_{\ensuremath{\it trigg}}^{e^+ e^-} \,=\, 0.9 \pm 0.1$.

\item At least 30~GeV energy deposited in one or both of the ZDCs is required 
to select Au+Au events with forward neutron emission ($Xn$) from the (single 
or double) Au$^\star$ decay. As has been discussed, this requirement leaves 
about 55\% of the coherent and $\approx$100\% of the incoherent $\jpsi$ 
events.

\end{enumerate}

The trigger efficiency is estimated to be $\varepsilon_{\ensuremath{\it 
trigg}}^{e^+ e^-} \,=\, 0.9 \pm 0.1$ from the ERT requirement. The 
inefficiencies due to requirements (1) and (3) were calculated and found to 
be negligible.

The total number of events collected by the UPC trigger was 8.5 M, of which 
6.7 M satisfied standard data quality assurance criteria. The useable event 
sample corresponds to an integrated luminosity $ {\mathcal L}_{int} = 141 \pm 
12 \;\mu\mbox{b}^{-1}$ computed from the minimum bias triggered events.

\section{Data Analysis}
\label{section:analysis}

Charged particle tracking in the PHENIX central arms is based on a 
combinatorial Hough transform in the track bend plane (perpendicular to the 
beam direction). The polar angle is determined from the position of the track 
in the PC outside the DC and the reconstructed position of the collision 
vertex~\cite{Mitchell:2002wu}. For central collisions, the collision vertex 
is reconstructed from timing information from the BBC and/or ZDC. This does 
not work for UPC events, which, by definition, do not have BBC coincidences 
and often do not have ZDC coincidences. The event vertex was instead 
reconstructed from the position of the PC hits and EMCal clusters associated 
with the tracks in the event. This gave an event vertex resolution in the 
longitudinal direction of 1~cm. Track momenta are measured with a resolution 
$\delta p/p \approx$ 0.7\%~$\oplus$~1.0\%~$p$[GeV/c] in minimum bias Au+Au 
nuclear collisions~\cite{Adler:2003rc}. Only a negligible reduction in the 
resolution is expected in this analysis because of the different vertex 
resolution.

The following global cuts were applied to enhance the sample of genuine 
$\gamma$-induced events:
\begin{enumerate}

\item A standard offline vertex cut $|vtx_{z}| <$ 30 cm was required to 
select collisions well centered in the fiducial area of the central detectors 
and to avoid tracks close to the magnet poles.

\item Only events with two charged particles were analyzed. This is a 
restrictive criterion imposed to cleanly select ``exclusive'' processes 
characterised by only two isolated particles (electrons) in the final state. 
It allows to suppress the contamination of non-UPC (mainly beam-gas and 
peripheral nuclear) reactions that fired the UPC trigger, whereas the signal 
loss is small (less than $5\%$).

\end{enumerate}

Unlike the $\jpsi\rightarrow e^+e^-$ analyses in nuclear Au+Au 
reactions~\cite{Adler:2003rc,Adare:2006ns} which have to deal with large 
particle multiplicities, we did not need to apply very strict electron 
identification cuts in the clean UPC environment. Instead, the following 
RICH- and EMCal-based offline cuts were used:

\begin{enumerate}

\item RICH multiplicity $n_0\geq$2 selects $e^\pm$ which fire 2 or more tubes 
around the track within the nominal ring radius.

\item Candidate tracks with an associated EMCal cluster with dead or noisy 
towers within a 2$\times$2 tile are excluded.

\item At least one of the tracks in the pair is required to pass an EMCal 
cluster energy cut ($E_1 >$~1~GeV $||$ $E_2 >$~1~GeV) to select candidate 
$e^\pm$ in the plateau region above the turn-on curve of the ERT trigger 
(which has a 0.8~GeV threshold).

\end{enumerate}

\noindent
Beyond those global or single-track cuts, an additional ``coherent'' 
identification cut was applied by selecting only those $e^+e^-$ candidates 
detected in opposite arms. Such a cut aims at reducing the high-$p_{\rm T}$ pairs 
while improving the detection of the low-$p_{\rm T}$ pairs expected for 
$\gaga,\,\gA$ production. Nevertheless, after all the previous cuts were 
applied the influence of this selection is found to be small; there is only 
one event in which the $e^+$ and $e^-$ are in the same arm and have 
$m_{e^+e^-}>2$~GeV/c$^2$.

Finally, $\jpsi$ were reconstructed by invariant mass analysis of the 
measured $e^+ e^-$ pairs. There was no remaining like-sign background after 
the aforementioned analysis cuts.

The cross sections were obtained after correcting the raw number of signal 
counts for the geometrical acceptance of our detector system, and the 
efficiency losses introduced by the previously described analysis cuts. 
Acceptance and efficiency corrections were obtained using the PHENIX {\sc 
geant3}~\cite{geant} simulation package with input distributions from the 
{\sc starlight} Monte Carlo (MC), based on the models presented 
in~\cite{Baltz:2002pp,Klein:1999qj,Nystrand:2004vn}. The measured $\gamma + p 
\rightarrow V+p$ cross sections from HERA and fixed target experiments with 
lepton beams are used as input to the models. {\sc starlight} well reproduces 
the existing $d^3N/dyd\phi dp_{\rm T}$ distribution of coherent $\rho$ production 
in UPC Au+Au events measured at RHIC by 
STAR~\cite{Adler:2002sc,Abelev:2007nb}. Helicity conservation is assumed in 
the model, and the angular distribution of the decay products ($\jpsi 
\rightarrow e^+ e^-$) is given by $dN/d\cos(\theta) \propto 1 + 
\cos^2(\theta)$ in the $\jpsi$ center-of-mass. The angular distribution is 
different from that for $\rho$ production followed by the decay $\rho 
\rightarrow \pi^+ \pi^-$, because of the different spin of the daughters, as 
well as from the angular distribution in two-photon interactions $\gamma + 
\gamma \rightarrow e^+ e^-$. We generated 5$\cdot$10$^4$ coherent $\jpsi$ and 
8$\cdot$10$^6$ coherent high-mass $e^+e^-$ pairs ($m_{e^+e^-}>1$~GeV/$c^2$) 
in Au+Au collisions accompanied by forward neutron emission.  The simulated 
events were passed through the same reconstruction programme as the real 
data.

Table~\ref{tab:efficiencies} summarises the $\jpsi$ and dielectron acceptance 
and efficiency correction factors obtained from our simulation studies. For 
instance, for $\jpsi$ photoproduction the correction is $1/(2.49 \pm 0.25) 
\%$, of which the experimental acceptance to detect the decay electron pair 
is about $5\%$ (for $\jpsi$ produced at $|y|<0.35$). 
In the $\gamma \gamma \rightarrow e^+e^-$ sample, most of the
electrons/positrons are emitted at very forward angles. The fraction of events
with $|y_{\rm pair}|<0.35$ and 
$2.0<m_{e^+e^-}<2.8~{\rm GeV}/c^2$, 
where both the electron and
positron are within $|\eta|<0.35$ is 1.10\%. 
The corresponding numbers for
$2.0<m_{e^+e^-}<2.3~{\rm GeV}/c^2$ and 
$2.3<m_{e^+e^-}<2.8~{\rm GeV}/c^2$ are 1.11\% and 1.08\%, respectively. 
The acceptance and efficiency corrections have a systematic uncertainty of 10\%
resulting from the accuracy of the simulation to describe the detector, the
electron identification parameters, and the event vertex position resolution.

%|%%%%%%%%%%%%%%%%%%%%%%%%%%%%%%%%%%%%%%%%%%%%%%%%%%% Table I
\begin{table}[tbh]
  \caption{
    \label{tab:efficiencies}
Coherent $\jpsi$ and $e^+e^-$ (continuum) acceptance and efficiency for 
$|y_\mathrm{pair}|<0.35$ as a function of invariant mass range. 
The last line shows the trigger efficiency.
  }
  \begin{center}
    \begin{tabular}{cc}
      \noalign{\smallskip}\hline\hline 
      $m_{e^+e^-}$ [GeV/c$^2$] & $\ensuremath{{\it Acc}}$ $\times$ $\varepsilon$ 
      \\  \hline\hline 
      $\jpsi$  & $ ( 2.49 \pm 0.25) \cdot 10^{-2} $
      \\
      $e^+e^- $ \mbox{[2.0,2.8]} & $ ( 2.24 \pm 0.22) \cdot 10^{-3} $ 
      \\
      $e^+e^- $ \mbox{[2.0,2.3]} & $ ( 2.16 \pm 0.22 ) \cdot 10^{-3} $ 
      \\
      $e^+e^- $ \mbox{[2.3,2.8]} & $ ( 2.33 \pm 0.23 ) \cdot 10^{-3} $ 
      \\ \hline 
      $\epsilon_{trigg}$           & $0.9 \pm 0.1$
      \\  \hline\hline 
    \end{tabular}
  \end{center}
\end{table}

\section{Results and Discussion}
\label{section:results}

\begin{figure}[tb]
\begin{center}
\includegraphics[width=0.56\linewidth]{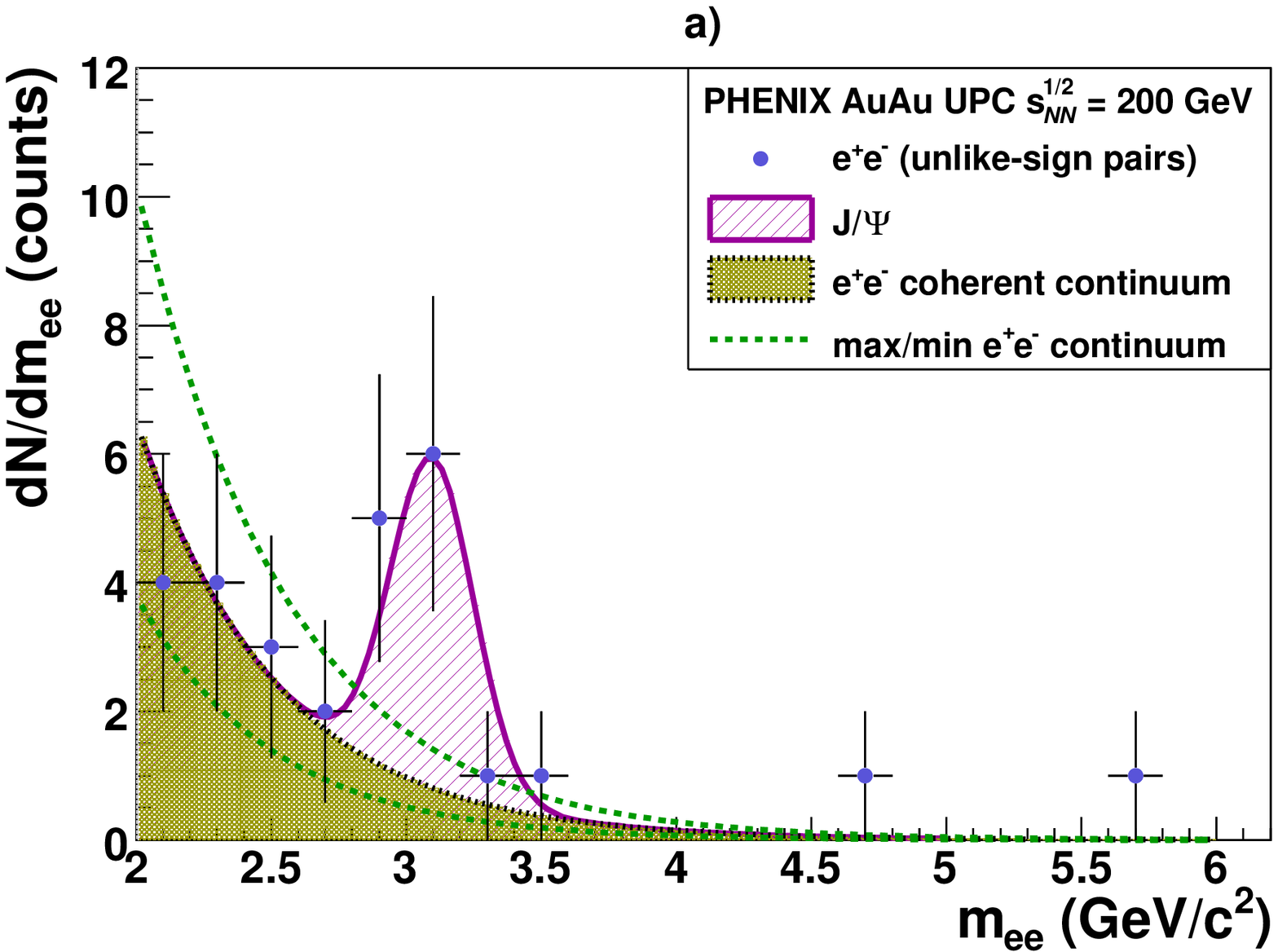} 
\includegraphics[width=0.43\linewidth]{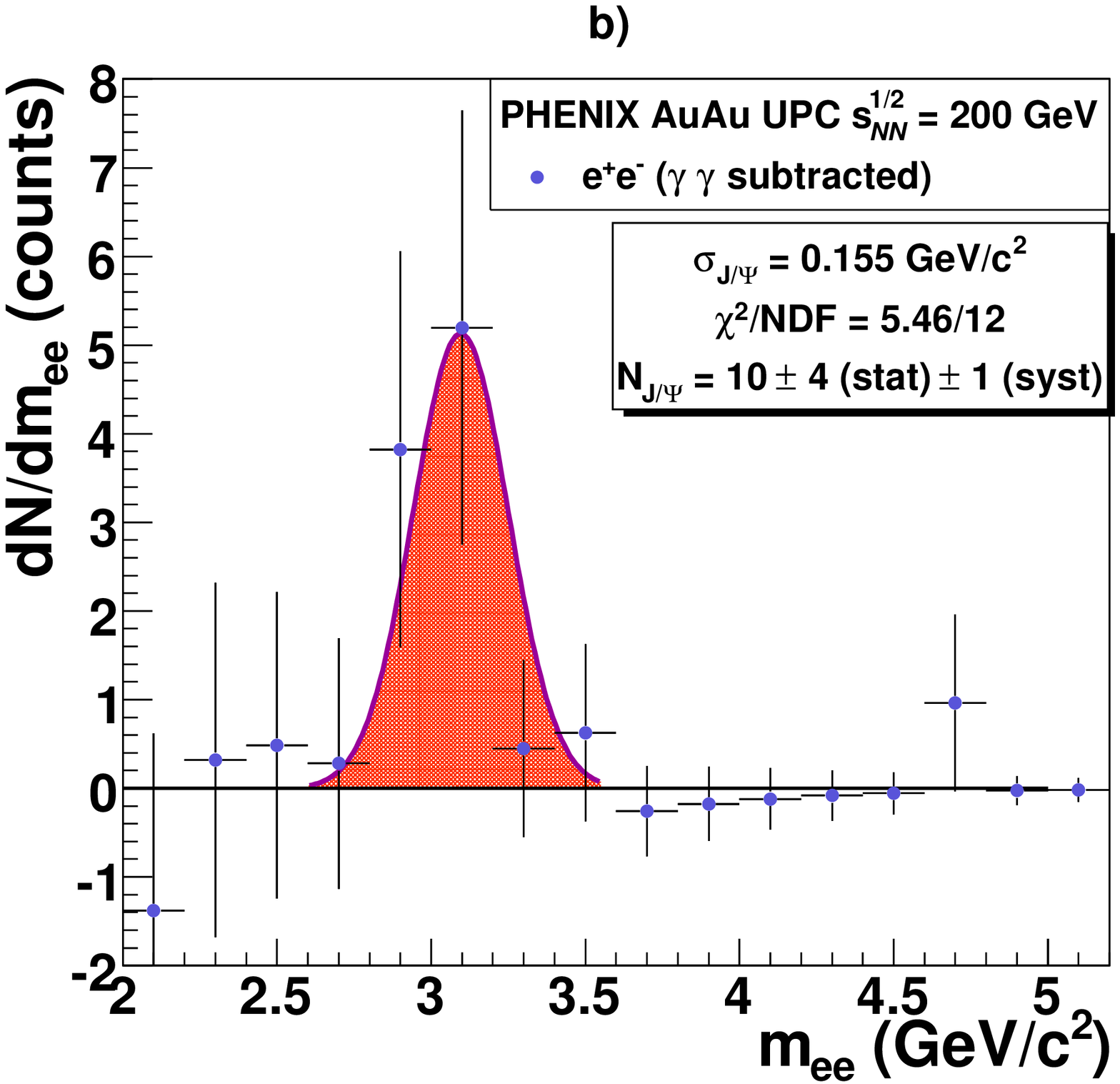} 
\end{center}
\caption{Left: 
  a) Invariant mass distribution of $e^+e^-$ pairs fitted to the 
combination of a dielectron continuum (exponential distribution) and a $\jpsi$ 
(Gaussian) signal. The two additional dashed curves indicate the fit
results with the maximum and minimum continuum contributions considered in this analysis (see text). 
 b) $\jpsi$ invariant mass distribution after subtracting the fitted dielectron continuum signal in a).
}
\label{fig:minv_ee_jpsi}
\end{figure}

The measured $e^+ e^-$ invariant mass distribution for the sample is shown in 
\fig{fig:minv_ee_jpsi}~a). The amount of background can be estimated from the 
number of like-sign events (i.e. events where two electrons or two positrons 
are reconstructed). We find no like-sign pairs for 
$m_{e^{\pm}e^{\pm}}>2$~GeV/c$^2$, compared with 28 events with an $e^+ e^-$ 
pair with $m_{e^+e^-}>2$~GeV/c$^2$. The shape is consistent with the expected 
contribution from the two processes in \fig{fig:diag_gg_gA}: a continuum 
distribution corresponding to two-photon production of $e^+ e^-$ pairs and a 
contribution from $\jpsi \rightarrow e^+ e^-$. Since the offline cuts ($E_1 
>$~1~GeV $||$ $E_2 >$~1~GeV) cause a sharp drop in the efficiency for 
$m_{e^+e^-} <$~2~GeV$/c^2$, we include only pairs with $m_{e^+e^-} 
\geq$~2~GeV$/c^2$ in the analysis.

The invariant mass distribution is fitted with a continuum (exponential) 
curve combined with a Gaussian function at the $\jpsi$ peak, as shown by the 
solid curve in \fig{fig:minv_ee_jpsi}~a). Simulations based on events 
generated by the {\sc starlight} MC (see last paragraphs of 
Section~\ref{section:analysis}) processed through {\sc geant} have shown that 
the shape of the measured continuum contribution is well described by an 
exponential function $dN/dm_{e^+ e^-} = A \cdot e^{ c \, m_{e^+ e^-}}$. Those 
simulations allow us to fix the exponential slope parameter to $c=-1.9 \pm 
0.1$~GeV$^{-1}$c$^2$.  The combined data fit is done with three free 
parameters: the exponential normalisation ($A$), the $\jpsi$ yield and the 
$\jpsi$ peak width (the Gaussian peak position has been fixed at the known 
$\jpsi$ mass of $m_{\jpsi}$~=~3.097~GeV/c$^2$~\cite{pdg}). 
\fig{fig:minv_ee_jpsi}~b) shows the resulting invariant mass distribution 
obtained by subtracting the fitted exponential curve of the dielectron 
continuum from the total experimental $e^+e^-$ pairs distribution. There is a 
clear $\jpsi$ peak, the width of which ($\sigma_{\jpsi} \sim 155$~MeV/$c^2$) 
is consistent with the $\jpsi$ width from our full MC.

%|%%%%%%%%%%%%%%%%%%%%%%%%%%%%%%%%%%%%%%%%%%%%%%%%%%%%%%%%%%% Fig_2

The $\jpsi$ and continuum yields and the corresponding statistical errors are 
calculated from the fit. The dashed curves in \fig{fig:minv_ee_jpsi}~a) show 
the maximum and minimum $e^+e^-$ continuum contributions considered, 
including both the statistical and systematical uncertainties. The systematic 
uncertainties were determined varying the dielectron continuum subtraction 
method using a power-law form instead of an exponential function and by 
modifying the corresponding fitted slope parameters by $\pm 3 \sigma$. The 
propagated uncertainty of the extracted yields was estimated to be one count 
in both cases. The total number of $\jpsi$'s is: $N_{\jpsi} \,=\, 9.9 \pm 
4.1~\mbox{(stat)} \pm 1.0~\mbox{(syst)}$, and the number of $e^+ e^-$ 
continuum pairs for $m_{e^+e^-}\in~\mbox{[2.0,2.8]}$~GeV/$c^2$ is: $N_{e^+ 
e^-} \,=\, 13.7 \pm 3.7~\mbox{(stat)} \pm 1.0~\mbox{(syst)}$. 
\tab{tab:yields} shows the obtained results per invariant mass range.

%|%%%%%%%%%%%%%%%%%%%%%%%%%%%%%%%%%%%%%%%%%%%%%%%%%%%%%%%%%%% Table_II
\begin{table}[tb]
  \caption{
    \label{tab:yields}
    $\jpsi \rightarrow e^+e^-$ and $e^+e^-$ continuum yields obtained
    from the fit of the data to an exponential plus Gaussian function per
    invariant mass range. Systematic errors are obtained as described in the text.
  }
  \begin{center}
    \begin{tabular}{cc}
      \noalign{\smallskip} \hline\hline 
      $m_{e^+e^-}$ [GeV/c$^2$] & Yield
      \\  \hline\hline 
      $\jpsi$ & $N_{\jpsi} \,=\, 9.9 \pm 4.1~\mbox{(stat)} \pm 1.0~\mbox{(syst)}$
      \\
      $e^+e^- $ \mbox{[2.0,2.8]} & $N_{e^+ e^-} \,=\, 13.7 \pm 3.7~\mbox{(stat)} \pm 1.0~\mbox{(syst)}$
      \\
      $e^+e^- $ \mbox{[2.0,2.3]} & $N_{e^+ e^-} \,=\, 7.4 \pm 2.7~\mbox{(stat)} \pm 1.0~\mbox{(syst)}$
      \\
      $e^+e^- $ \mbox{[2.3,2.8]} & $N_{e^+ e^-} \,=\, 6.2 \pm 2.5~\mbox{(stat)} \pm 1.0~\mbox{(syst)}$
      \\  \hline\hline 
    \end{tabular}
  \end{center}
\end{table}

\begin{figure}[tbh] 
\begin{center}
\includegraphics[width=0.49\linewidth]{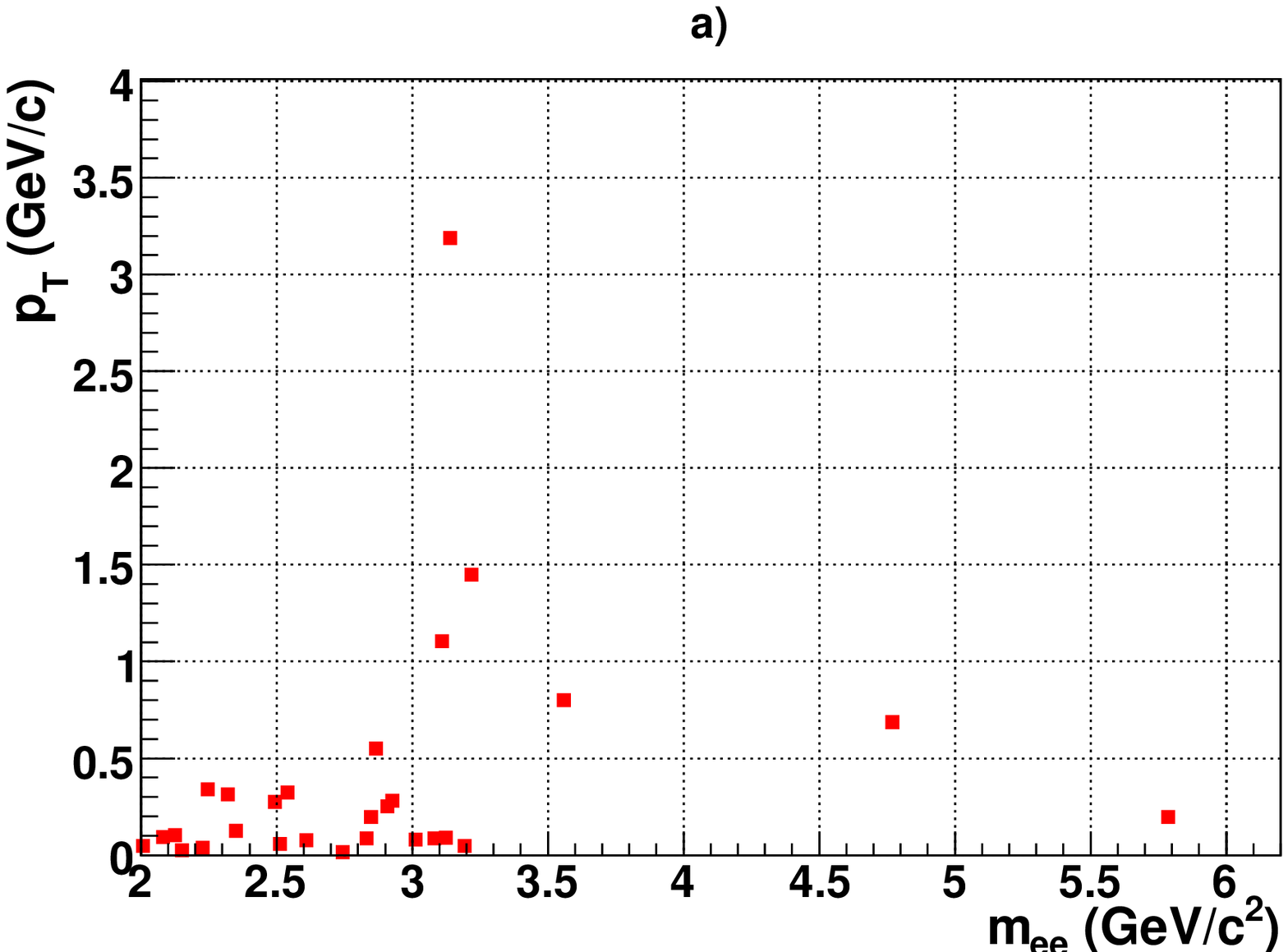} 
\includegraphics[width=0.49\linewidth]{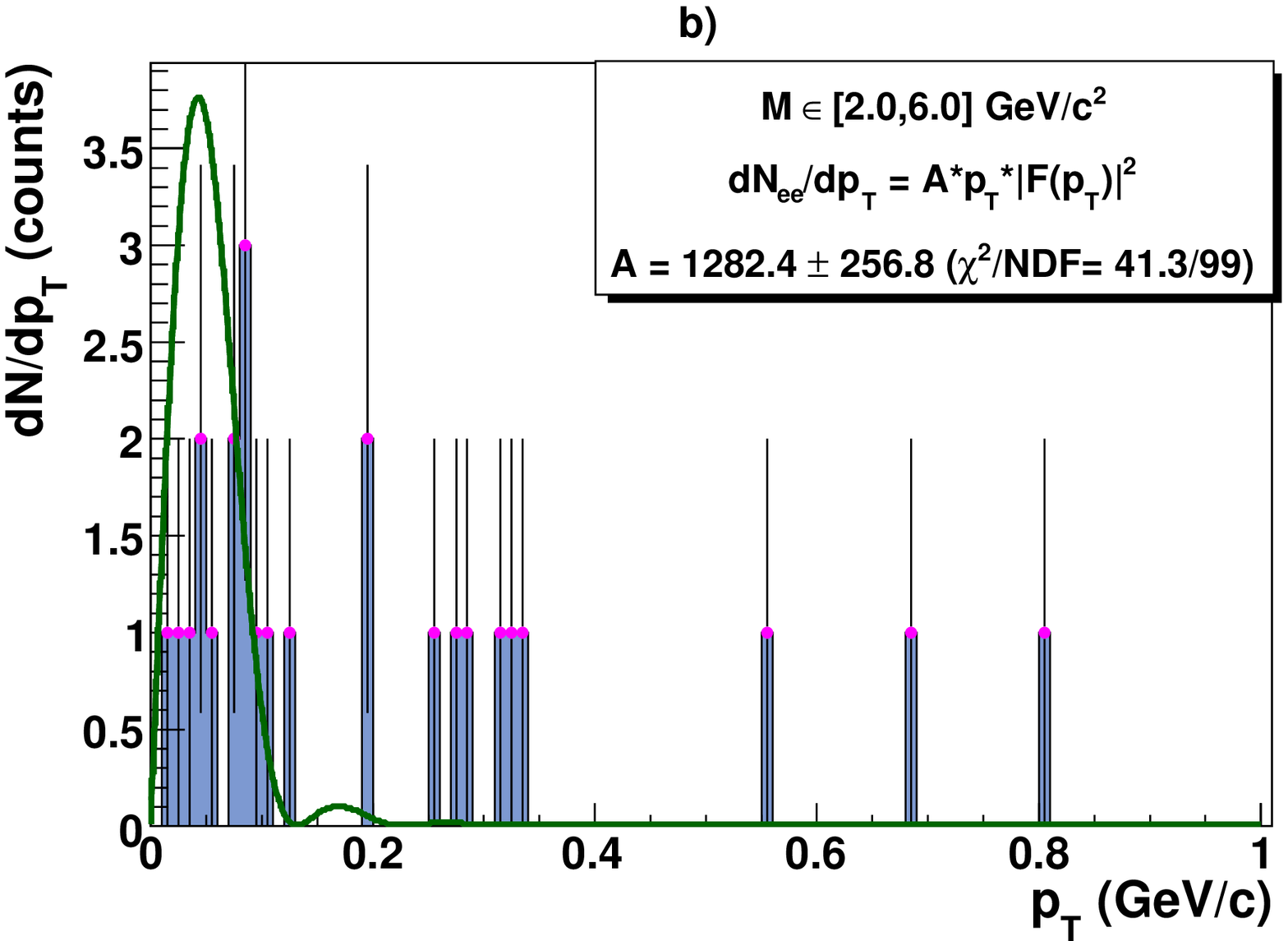}
\includegraphics[width=0.49\linewidth]{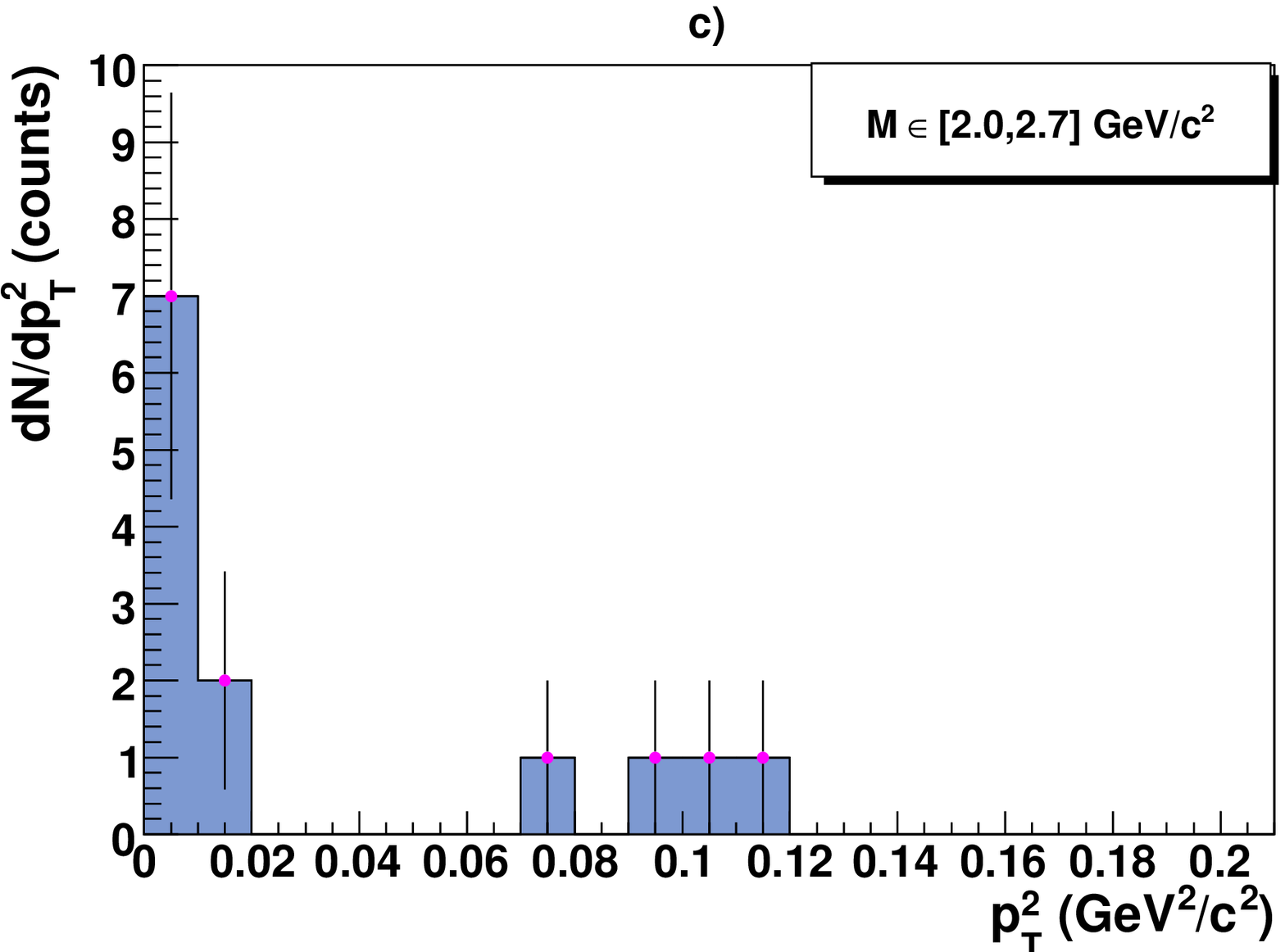}
\includegraphics[width=0.49\linewidth]{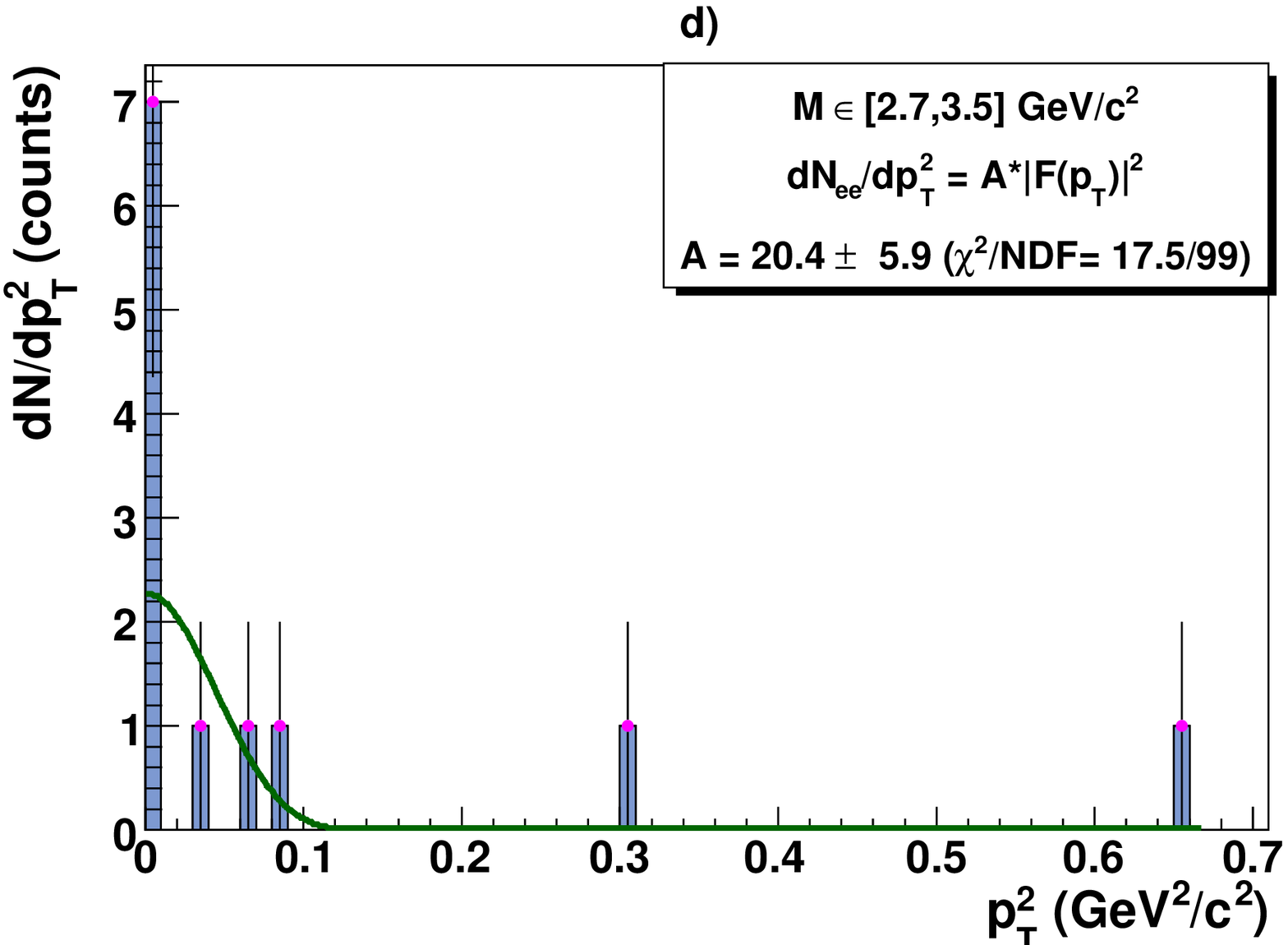}
\end{center}
  \caption{
    Top: (a) Scatter plot of $e^+ e^-$ $m_{e^+e^-}$ vs. pair $p_{\rm T}$. (b)
    $dN/dp_{\rm T}$ distribution of the pairs with $m_{e^+e^-}>2$~GeV/c$^2$ fitted to
    the Au nuclear form factor, Eq.~(\ref{eq:F_Au}). 
    Bottom: $dN/d p_{\rm T}^2$ distribution of pairs with (c) $m_{e^+e^-}\in
    \mbox{[2.0,2.7]}$~GeV/c$^2$ and (d) 
    $m_{e^+e^-}\in \mbox{[2.7,3.5]}$~GeV/c$^2$, fitted to
    the expected Au nuclear form factor. Note the difference in scale 
    on the $x$-axis in the lower two plots. 
  }
  \label{fig:minv_ee_pt}
\end{figure}

\fig{fig:minv_ee_pt}~a) shows a scatter plot of invariant mass $m_{e^+e^-}$ 
vs. pair $p_{\rm T}$. From the plot, it is clear that most of the pairs outside the 
$\jpsi$ peak originate in coherent processes with very low pair transverse 
momenta ($p_{\rm T} \lesssim 100$~MeV/c), as expected for two-photon interactions. 
For events with $m_{e^+e^-}$ around the $\jpsi$ mass, however, there are a 
few counts at larger $p_{\rm T}$ values which cannot be ascribed to background 
events, since there are no like-sign pairs above~2~GeV/c$^2$. A purely 
coherent production -- corresponding to events where the fields couple 
coherently to all nucleons and the nucleus remains in its ground state 
($\gamma + A \rightarrow V + A$) -- would yield $p_{\rm T}\lesssim$~200~MeV/c after 
reconstruction. On the other hand, incoherent production ($\gamma + A 
\rightarrow V + X$) -- dominated by the quasi-elastic vector meson production 
off one nucleon inside the nucleus, $\gamma + N \rightarrow V + N$ -- results 
in much larger $p_{\rm T}$ for the photoproduced $\jpsi$~\cite{Strikman:2005ze}. 
The cross sections for coherent and incoherent $\jpsi$ photoproduction in 
UPCs at RHIC are expected to be of the same order~\cite{Strikman:2005ze}. We 
discuss below whether our data confirm such a prediction.

%%%%%%%%%%%%%%%%%%%%%%%%%%%%%%%%%%%%%%%%%%%%%%%%%%%%%%%%%%%% Fig_3

The transverse momentum distribution of the events with 
$m_{e^+e^-}>2$~GeV/c$^2$ is shown in \fig{fig:minv_ee_pt} b). For clarity, 
only points below $p_{\rm T}<1$~GeV/c are drawn. The $p_{\rm T}$ is here the magnitude of 
the vector sum of the $\vec{p_{\rm T}}$ of the electron and positron. One sees a 
clear enhancement of events with very low transverse momenta, consistent with 
coherent production. The squared form-factor of a gold nucleus,
\begin{equation}
| \, F_{Au}(p_{\rm T}^2) \, |^2 = \left| \; 3 \; \frac{\sin(R p_{\rm T}) - R p_{\rm T} \cos( R p_{\rm T})}{(R p_{\rm T})^3 \left(1 + (a p_{\rm T})^2\right)} \; \right|^2,
\label{eq:F_Au}
\end{equation}
is shown for comparison. Here, $R = 6.7$~fm is the gold radius, and 
$a=0.7$~fm represents the diffuseness of the nuclear 
surface~\cite{Davies:1976zz}. The magnitude of the form-factor is a free 
parameter fitted to reproduce the spectra. \fig{fig:minv_ee_pt} also presents 
the corresponding distribution expressed in terms of the squared momentum 
transfer from the target nucleus, $p_{\rm T}^2\approx -t$, for events with 
$m_{e^+e^-}$ corresponding to the dielectron continuum $m_{e^+e^-} 
\in$~\mbox{[2.0,2.8]}~GeV/c$^2$ (\fig{fig:minv_ee_pt} c) and below the 
$\jpsi$--peak, $m_{e^+e^-} \in~\mbox{[2.7,3.5]}$~GeV/c$^2$ 
(\fig{fig:minv_ee_pt} d).

The extracted yields of $\jpsi$ and $e^+ e^-$ are used to calculate the final 
cross section for photoproduction at midrapidity in ultra-peripheral Au+Au 
collisions accompanied by forward neutron emission. For dielectrons at 
midrapidity ($y$ is the rapidity of the pair) the double differential cross 
section is:

\begin{eqnarray}
 \left .\frac{d^2 \sigma_{e^+ e^- + Xn}}{d y \, d
     m_{e^+e^-}}\right|_{|y|<0.35, \, \Delta m_{e^+e^-}} = 
  \frac{N_{e^+ e^-}}{ \ensuremath{{\it Acc}} \cdot \varepsilon \cdot
    \varepsilon_{trigg} \cdot \mathcal{L}_{int} } \cdot
  \frac{1}{\Delta y} \cdot \frac{1}{\Delta m_{e^+e^-}} \qquad \\
  =  86  \pm  23\, {\rm(stat)} \pm 16 \, {\rm(syst)} \; \mu {\rm b/}({\rm
    GeV/c}^2) \; \; {\rm for} \; m_{e^+e^-}\in \mbox{[2.0,2.8] GeV/c}^2.
  \nonumber
\end{eqnarray} 

For $\jpsi$ at midrapidity the differential cross section is:

\begin{eqnarray}
\left .\frac{d\sigma_{\jpsi + Xn}}{dy}\right|_{|y|<0.35} =  
\frac{1}{\ensuremath{{\it BR}}}\cdot\frac{N_{\jpsi}}{\ensuremath{{\it
      Acc}}\cdot \varepsilon \cdot
\varepsilon_{\ensuremath{\it trigg}}\cdot {\mathcal L}_{int}}\cdot\frac{1}{\Delta y} \\
= 76 \pm 31 \, {\rm(stat)} \pm 15 \, {\rm(syst)} \; \mu {\rm b}.
\nonumber
\end{eqnarray}
\noindent
The correction factors (and corresponding uncertainties) are quoted in 
Table~\ref{tab:efficiencies} as described in previous sections, and BR = 
5.94\% is the known $\jpsi$ dielectron branching ratio~\cite{pdg}. 
\tab{tab:xs_results} summarises the measured cross sections per invariant 
mass interval.

\begin{table}[!htbp]
  \caption{
    \label{tab:xs_results}
Measured $\jpsi$ and $e^+ e^-$ continuum photoproduction cross sections at 
midrapidity in ultra-peripheral Au+Au collisions (accompanied with forward 
neutron emission) at $\sqrtsnn$~=~200~GeV. The rightmost column in the lower 
part shows the {\sc starlight} predictions~\cite{Nystrand:2004vn}.
  }
  \begin{center}
    \begin{tabular}{ccc}
      \noalign{\smallskip} \hline\hline
        & \multicolumn{2}{c}{$d\sigma/ dy|_{y=0}$ [$\mu$b]} 
      \\  \hline
      $J/\psi$ & \multicolumn{2}{c}{$ 76 \pm 31 \, {\rm(stat)} \pm 15 \, {\rm(syst)} $}
      \\  \noalign{\smallskip} \hline\hline 
      $m_{e^+e^-}$ [GeV/c$^2$] & \multicolumn{2}{c}{$d^2\sigma/dm_{e^+e^-} dy|_{y=0}$ [$\mu$b/(GeV/c$^2$)]}  
      \\  
                              & data                                       & {\sc starlight}
      \\  \hline
      $e^+e^-$ continuum [2.0,2.8] & $  86 \pm  23\, {\rm(stat)} \pm 16 \, {\rm(syst)} $ & $90$ \\
      $e^+e^-$ continuum [2.0,2.3] & $ 129 \pm  47 \, {\rm(stat)} \pm  28 \, {\rm(syst)}$ & $138$ \\  
      $e^+e^-$ continuum [2.3,2.8] & $  60 \pm  24 \, {\rm(stat)} \pm  14 \, {\rm(syst)} $ & $61$ \\ 
      \hline\hline 
    \end{tabular}
  \end{center}
\end{table}

The measured dielectron cross sections at midrapidity are in very good 
agreement with the {\sc starlight} predictions for coherent dielectron 
photoproduction (rightmost column of 
Table~\ref{tab:xs_results})~\cite{Nystrand:2004vn}. Exclusive dilepton 
production in {\sc starlight} is calculated combining the two equivalent 
(Weizs\"acker-Williams) photon fluxes from each ion with the Breit-Wheeler 
formula for $\gamma\gamma \rightarrow l^+l^-$. The agreement between {\sc 
starlight} and other leading order calculations~\cite{Baur:2007zz} is good as 
long as the pair invariant mass is not too low. A recent calculation has 
found that higher order terms suppress the $e^+e^-$ cross section by 29\% in 
the invariant mass range 140~$< m_{e^+e^-} <$~165 
MeV/c$^2$~\cite{Baltz:2007gs}. A reduction of the same magnitude in the 
invariant mass range considered here, 2.0~$< m_{e^+e^-} <$~2.8~GeV/c$^2$, 
would still be in agreement with our measurement.

\begin{figure}[tbh]
\begin{center}
\includegraphics[width=0.49\columnwidth]{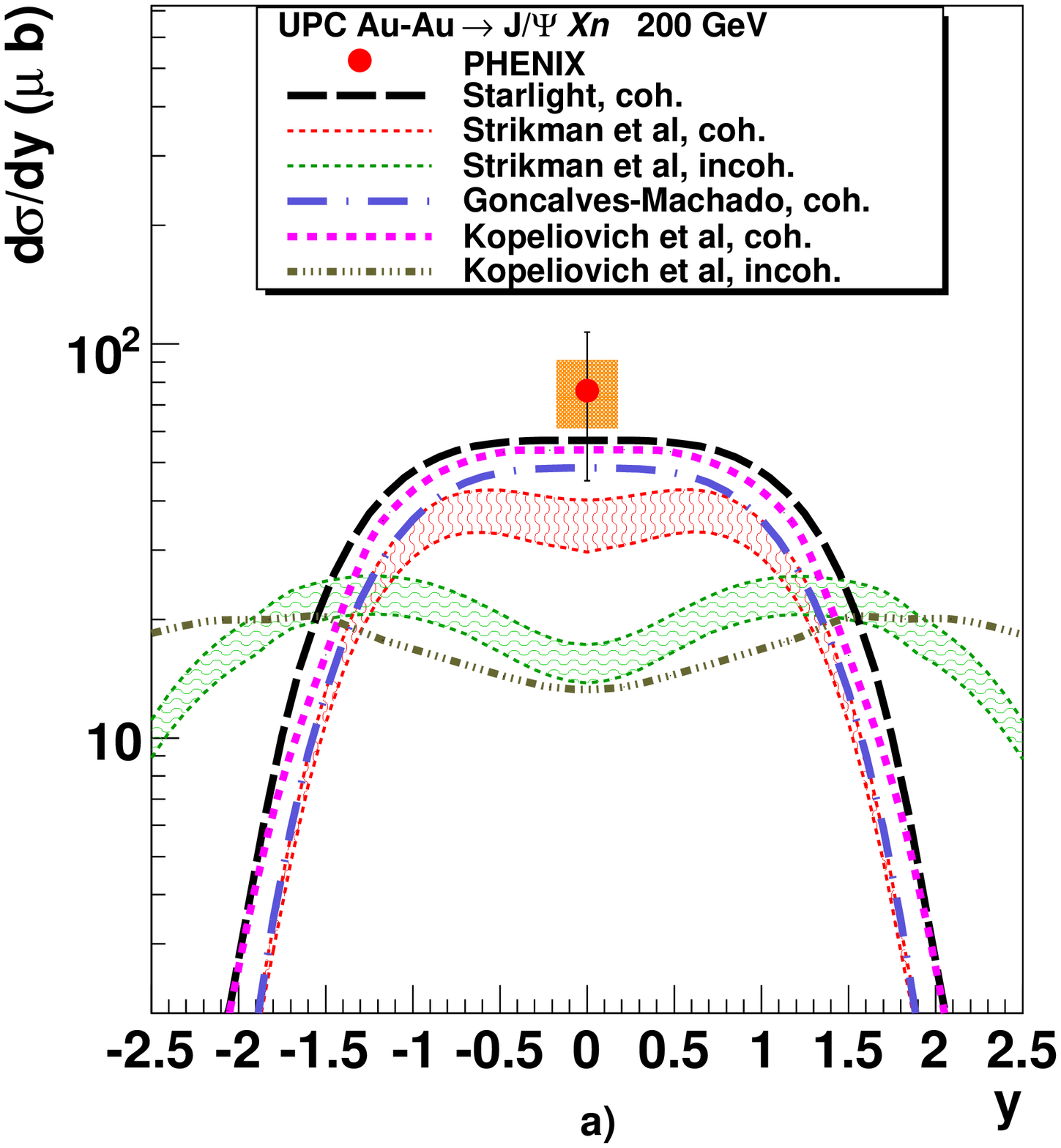}
\includegraphics[width=0.49\columnwidth]{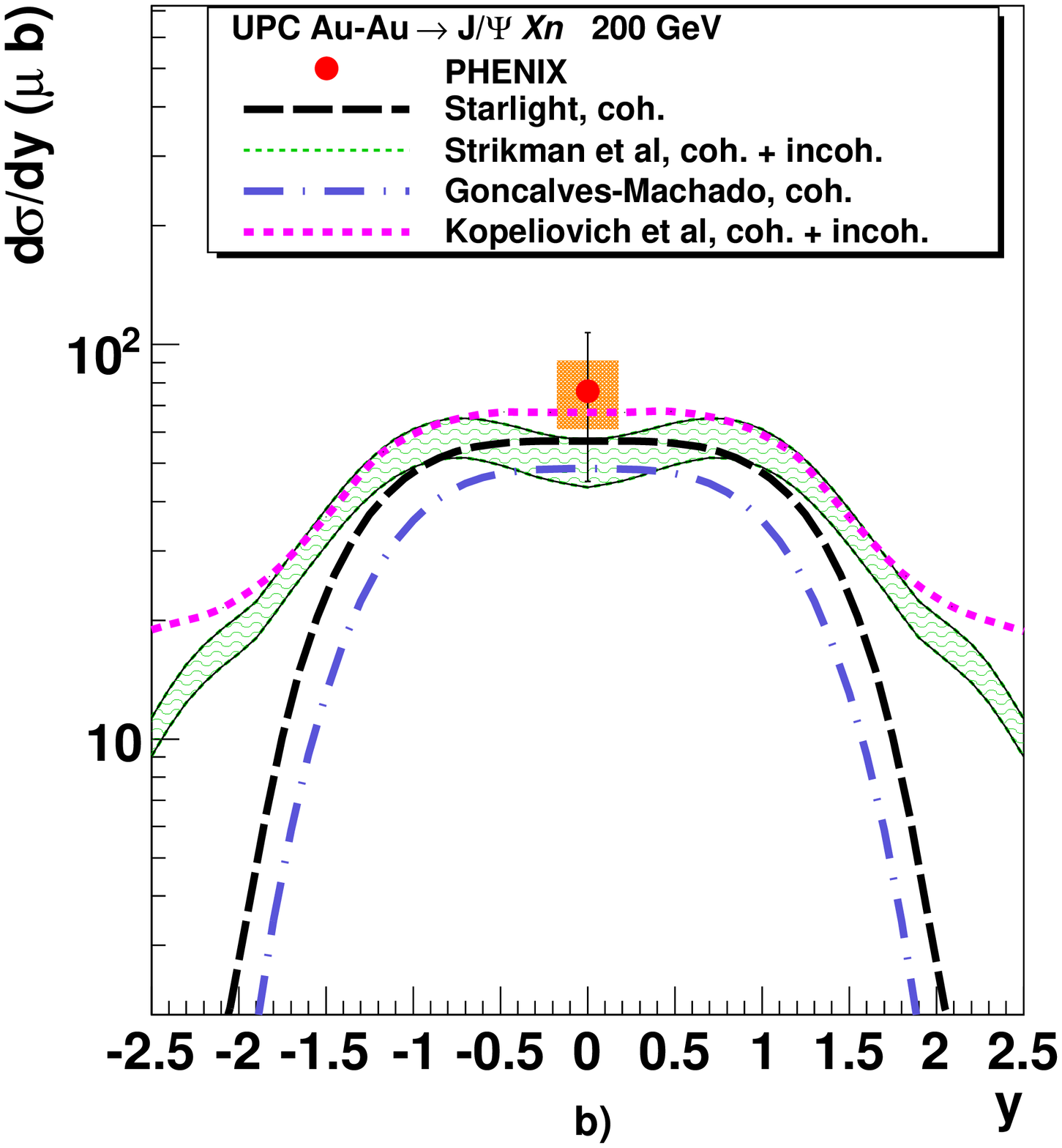}
\end{center}
\caption{
Cross section of $\jpsi + Xn$ production at midrapidity in UPC Au+Au 
collisions at $\sqrtsnn$~=~200~GeV compared to theoretical 
calculations:  
{\sc starlight}~\protect\cite{Baltz:2002pp,Klein:1999qj,Nystrand:2004vn},
Strikman, et al.~\protect\cite{Strikman:2005ze}, 
Goncalves-Machado~\protect\cite{Goncalves:2007qu}, and 
Kopeliovich, et al.,~\protect\cite{Ivanov:2007ms}.
The error bar (box) shows the statistical (systematical) uncertainties of the 
measurement. When available, the theoretical calculations for the coherent 
and incoherent components are shown separately in a), and summed up in b).}
\label{fig:dNdy_vs_model}
\end{figure}

The final $\jpsi +Xn$ cross section is compared to the theoretical 
predictions computed in 
references~\cite{Baltz:2002pp,Strikman:2005ze,Klein:1999qj,Ivanov:2007ms,Nystrand:2004vn,Goncalves:2007qu} 
in \fig{fig:dNdy_vs_model}. The rapidity distributions of Strikman {\it et 
al.}~\cite{Strikman:2005ze} and Kopeliovich {\it et al.}~\cite{Ivanov:2007ms} 
have been scaled down according to~\cite{Baltz:2002pp} to account for the 
reduction of the yield expected when requiring coincident forward neutron 
emission ($Xn$). The scaling has been applied as a function of rapidity with 
the integrated cross section being 55\% of the original one. The band covered 
by the Strikman {\it et al.} predictions includes the $\jpsi$ cross sections 
with and without gluon shadowing (as implemented in the 
Frankfurt-Guzey-Strikman, FGS, prescription~\cite{Frankfurt:2003zd}). 
Strikman and Kopeliovich predictions for the coherent and incoherent 
photoproduction cross sections are drawn separately in 
\fig{fig:dNdy_vs_model}~a) and summed up in \fig{fig:dNdy_vs_model}~b). {\sc 
starlight}~\cite{Nystrand:2004vn} and 
Gon\c{c}alves-Machado~\cite{Goncalves:2007qu} calculations only evaluate the 
coherent contribution.

As mentioned above, the measured pair $p_{\rm T}$ distributions suggest coherent 
$\jpsi$ photoproduction ($\gamma + A \rightarrow \jpsi +X$) and a possible 
additional incoherent ($\gamma + N \rightarrow \jpsi + X$) contribution at 
higher $p_{\rm T}$. To give an indicative estimate of the size of the incoherent 
contribution, we can assume that it corresponds to the counts in the $\jpsi$ 
mass window with $p_{\rm T}^2 > 0.1~(0.05)$~GeV$^2$/c$^2$. This corresponds to 
about $4~(6)$ counts, which amounts to a contribution of about $40~(60)\%$ of 
the total $\jpsi$ production, compatible with the theoretical 
calculations~\cite{Strikman:2005ze}. The limited data statistics prevents us 
from separating in a more quantitative way the two components. Note that 
although the acceptance correction for the $\jpsi$ was calculated using a 
Monte Carlo which includes only the coherent component, the obtained 
correction is also a reasonable approximation for the incoherent component, 
provided that quasi-elastic scattering on a single nucleon, $\gamma + N 
\rightarrow V + N$, gives the main contribution. The polarisation of the 
vector meson will then be the same as for coherent production, and the 
reduction in acceptance because of the different $p_{\rm T}$ range will be of the 
order of $\sim$10-20\%. If the incoherent contribution to the total $\jpsi$ 
photoproduction was $40$\%, the coherent $\jpsi$ cross section would become 
$\sim 46~\mu$b.

Despite these uncertainties, the final $\jpsi$ cross section is in good 
agreement, within the (still large) statistical errors, with the theoretical 
values computed 
in~\cite{Baltz:2002pp,Strikman:2005ze,Klein:1999qj,Ivanov:2007ms,Nystrand:2004vn,Goncalves:2007qu} 
as shown in \fig{fig:dNdy_vs_model}. The current uncertainties unfortunately 
preclude any more detailed conclusion at this point regarding the two crucial 
ingredients of the models (nuclear gluon shadowing and $\jpsi$ nuclear 
absorption cross section). The statistical uncertainties can be improved with 
significantly higher Au+Au luminosities and a concurrent measurement of the 
$\jpsi$ in the dimuon decay channel in the more forward acceptances covered 
by the PHENIX muon spectrometers, as collected in year 2007 and expected in 
the future.

Finally, one can attempt to compare the obtained photonuclear $\jpsi$ cross 
sections to those from $e$-$p$ collisions at HERA by dividing the measured 
differential cross section ($d \sigma/dy$) with the (theoretical) equivalent 
photon spectrum ($dN_{\gamma}/d\omega$). At midrapidity: 
$\sigma_{_{\gamma\,A\to \jpsi\,A}} =(d\sigma_{_{AA\to 
\jpsi\,AA}}/dy)/(2\,dN_{\gamma}/d\omega)$, with $2\,N_{\gamma}$ = 6.7 (10.5) 
for the coherent (incoherent) spectrum at a photon-nucleon center-of-mass 
energy of $\mean{W_{\gamma\,p}}$~=~24~GeV. Assuming, for the sake of 
simplicity, a 50\%~--~50\% contribution from coherent and incoherent 
interactions in our total ultra-peripheral $\jpsi$ sample, the extracted 
photonuclear cross sections are: $\sigma(\gamma+Au \to \jpsi+Au)$~=~$5.7 \pm 
2.3 \, {\rm(stat)} \pm 1.2 \, {\rm(syst)} $~$\mu$b, and $\sigma(\gamma+Au \to 
\jpsi+X)$~=~$3.6 \pm 1.4 \, {\rm(stat)} \pm 0.7 \, {\rm(syst)} $~$\mu$b, 
respectively. A fit to the results from the H1 and ZEUS 
collaborations~\cite{Chekanov:2002xi,Aktas:2005xu} over their measured energy 
range gives $\sigma(\gamma+p \to \jpsi+p)$~=~30.5~$\pm$2.7~nb at 
$W_{\gamma\,p}$~=~24~GeV. Therefore, the ratios $\sigma(\gamma+Au \to 
\jpsi)/\sigma(\gamma+p\to \jpsi)$~=~186~$\pm$88, 118~$\pm$54 for the coherent 
and incoherent components (statistical and systematic errors assumed 
independent and added in quadrature) are consistent with a scaling of the 
photonuclear cross section with the number of nucleons in gold ($A$~=~179): 
$\sigma(\gamma+Au \to \jpsi)=A^\alpha\;\sigma(\gamma+p \to \jpsi)$ with 
$\alpha_{{\ensuremath{\it coh}}}$~=~1.01~$\pm$0.07, and 
$\alpha_{{\ensuremath{\it incoh}}}$~=~0.92~$\pm$0.08, 
respectively\footnote{Note, for comparison, that repeating the same exercise 
for the photoproduced $\rho$ in the STAR UPC 
measurement~\cite{Abelev:2007nb}, $\sigma(\gamma+Au \to \rho+Au)$~=~$530 \pm 
19 \, {\rm(stat)} \pm 57 \, {\rm(syst)} $~$\mu$b for $\mean{W_{\gamma 
N}}\sim$~12.5~GeV, and taking the experimentally-derived value 
$\sigma(\gamma+p\to\rho+p)$~=~9.88~$\mu$b from~\cite{Klein:1999qj}, yields 
$\alpha_{{\ensuremath{\it coh}}}$~=~0.75~$\pm$~0.02 closer to the 
$A^{2/3}$-scaling expected for soft particle production.}.

%%%%%%%%%%%%%%%%%%%%%%%%%%%%%%%%%%%%%%%%%%%%%%%%%%%%%%%%%%%% Fig_4

%%%%%%%%%%%%%%%%%%%%%%%%%%%%%%%%%%%%%%%%%%%%%%%%%%%%% Summary and Conclusions
\section{Summary and Conclusions}
\label{section:summary}

We have presented the first exclusive photoproduction of $\jpsi\to e^+e^-$ 
and high-mass $e^+e^-$ pairs ever measured in nucleus-nucleus (as well as 
hadron-hadron) interactions. The measurement has been 
carried out by the PHENIX experiment in ultra-peripheral Au+Au interactions 
at $\sqrtsnn$~=~200~GeV tagged by forward (ZDC) neutron detection from the 
(single or double) Au$^\star$ dissociation. Clear signals of $\jpsi$ and 
high mass dielectron continuum have been found in the data. We have 
observed 28 $e^+ e^-$ pairs in $m_{e^+e^-}\in$~\mbox{[2.0,6.0]~GeV/c}$^2$ 
with zero like-sign background. Their $p_{\rm T}$ spectrum is peaked at low 
$p_{\rm T}\approx$~90~MeV/$c$ as expected for coherent photoproduction with a 
realistic Au nuclear form factor.

The measured number of continuum $e^+ e^-$ events in the PHENIX acceptance 
for $m_{e^+e^-}\in$~\mbox{[2.0,2.8]~GeV/c}$^2$ is: $N(e^+ e^-)$ = 13.7 
$\pm$ 3.7 (stat) $\pm$ 1.0 (syst). After correcting for acceptance and 
efficiency losses and normalising by the measured luminosity, we obtain a 
cross section of $d^2 \sigma/dm_{e^+e^-} dy \,(e^+e^- + Xn)|_{y=0}$ = 86 
$\pm$ 23 (stat) $\pm$ 16 (syst) $\mu$b/(GeV/c$^2$), which is in good 
agreement with theoretical expectations for coherent exclusive dielectron 
production in photon-photon interactions.

The measured invariant mass distribution has a clear peak at the $\jpsi$ 
mass with an experimental width in good agreement with a full {\sc 
geant}-based simulation for UPC production and reconstruction in the PHENIX
detector.  The measured number of $\jpsi$ mesons in the PHENIX acceptance 
is: $N(\jpsi)$ = 9.9 $\pm$ 4.1 (stat) $\pm$ 1.0 (syst). 
The higher $p_{\rm T}$ distribution 
suggests an additional incoherent contribution to $\jpsi$ photoproduction 
in accordance with predictions~\cite{Strikman:2005ze}, but statistical 
limitations prevent a more quantitative estimate.  After 
correcting for acceptance and efficiency losses and normalising by the 
measured luminosity, the total $\jpsi$ photoproduction cross section is 
$d\sigma/dy \, (\jpsi + Xn)|_{y=0}$ = 76 $\pm$ 31 (stat) $\pm$ 15 (syst) 
$\mu$b, which is consistent (within uncertainties) with theoretical 
expectations. The low background in the present data sample shows that 
future higher luminosity runs with reduced experimental uncertainties of 
the measured cross sections will provide more quantitative information on 
the nuclear gluon distribution and $\jpsi$ absorption in cold nuclear 
matter at RHIC energies.

\section*{Acknowledgments}   % Run-4 and Run-5 short form for PRL

We thank the staff of the Collider-Accelerator and
Physics Departments at BNL for their vital contributions.
We acknowledge support from
the Department of Energy and NSF (U.S.A.),
MEXT and JSPS (Japan),
CNPq and FAPESP (Brazil),
NSFC (China),
MSMT (Czech Republic),
IN2P3/CNRS, and CEA (France),
BMBF, DAAD, and AvH (Germany),
OTKA (Hungary),
DAE (India),
ISF (Israel),
KRF and KOSEF (Korea),
MES, RAS, and FAAE (Russia),
VR and KAW (Sweden),
U.S. CRDF for the FSU,
US-Hungarian NSF-OTKA-MTA,
and US-Israel BSF.

\bibliographystyle{elsart-num}

%\bibliography{ppg081}

\end{document}